\documentstyle[12pt,aas2pp4,flushrt]{article}
\tightenlines
\def\kms{\relax \ifmmode {\,\rm km\,s}^{-1}\else \,km\,s$^{-1}$\fi}

\def\farcs{\hbox{$.\!\!^{\prime\prime}$}}

\def\secd#1.#2{ #1\farcs#2 }               

\def\td{$tD^{-1}$}

\def\mincir{\ \raise-2.truept\hbox{\rlap{\hbox{$\sim$}}\raise5.truept
    \hbox{$<$}\ }}
\def\magcir{\ \raise-2.truept\hbox{\rlap{\hbox{$\sim$}}\raise5.truept
    \hbox{$>$}\ }}
\def\gr{$^\circ$}

\def\nii{[N~{\sc ii}]}
\def\sii{[S~{\sc ii}]}
\def\oii{[O~{\sc ii}]}
\def\heii{He~{\sc ii}}
\def\oiii{[O~{\sc iii}]}
\def\ha{H$\alpha$}
\def\hb{H$\beta$}

\lefthead{Jets, knots and tails in planetary nebulae}
\righthead{Corradi et al.}

\begin{document}

\title{Jets, knots and tails in planetary nebulae:\\ NGC~3918,
K~1--2 and Wray 17-1\footnote{Based on observations with the 3.5-m~NTT and
1.54-m~Danish telescopes of the European Southern Observatory, and with the
NASA/ESA {\it Hubble Space Telescope}, obtained at the Space Telescope Science
Institute, which is operated by AURA for NASA under contract NAS5-26555.}}

\author{Romano L. M. Corradi}
        
\affil{Instituto de Astrof\'{\i}sica de Canarias, c. Via Lactea S/N, \\
	E--38200 La Laguna, Tenerife, Spain. \\e--mail: rcorradi@ll.iac.es}

\author{Mario Perinotto}

\affil{Dipartimento di Astronomia e Scienza dello Spazio, Universit\`a 
di Firenze, \\Largo E. Fermi 5, 50125 Firenze, Italy. 
\\e--mail: mariop@arcetri.astro.it}

\author{Eva Villaver, Antonio Mampaso, and Denise R. Gon\c calves}
        
\affil{Instituto de Astrof\'{\i}sica de Canarias, c. Via Lactea S/N,\\ E--38200
       La Laguna, Tenerife, Spain. \\e--mail: villaver@ll.iac.es,
       amr@ll.iac.es, denise@ll.iac.es}

\altaffiltext{1}{}

\begin{abstract} 
We analyze optical images and high-resolution, long-slit
spectra of three planetary nebulae which possess collimated, low-ionization
features.

NGC 3918 is composed of an inner, spindle-shaped shell mildly inclined with
respect to the plane of the sky.  Departing from the polar regions of this
shell, we find a two-sided jet expanding with velocities which increase linearly
with distance from 50 to 100~\kms.  The jet is probably coeval with the inner
shell (age $\approx 1000\cdot$$D$ y, where $D$ is the distance in kpc),
suggesting that its formation should be ascribed to the same dynamical processes
which also shaped the main nebula, and not to a more recent mass loss episode.
We discuss the formation of the aspherical shell and jet in the light of current
hydrodynamical and magnetohydrodynamical theories.

K 1-2 is a planetary nebula with a close binary nucleus which shows a collimated
string of knots embedded in a diffuse, elliptical shell.  The knots expand with
a velocity similar to that of the elliptical nebula ($\sim$25~\kms), except for
an extended tail located out of the main nebula, which linearly accelerates up
to $\sim 45$~\kms.  We estimate an inclination on the line of the sight of
$\sim$40\gr\ for the string of knots; once the orientation of the orbit is also
determined, this information will allow us to test the prediction of current
theories of the occurrence of polar jets from close binary systems.

Wray 17-1 has a complex morphology, showing two pairs of low-ionization
structures located in almost perpendicular directions from the central star, and
embedded in a large, diffuse nebula.  The two pairs show notable similarities
and differences, and their origin is very puzzling.

\end{abstract}


\section{Introduction}

Hydrodynamical modeling based on the interacting-winds theory (Kwok, Purton, \&
FitzGerald 1978; Kahn 1989) has proved to be quite successful in reproducing the
overall shapes of many planetary nebulae (PNe).  But PNe are also known to
contain a variety of small-scale structures in the form of low-ionization
patches, bullets, ansae, jets, point-symmetrical knots, filaments etc., whose
existence was noted in the past by several authors (e.g., Balick 1987; Schwarz
1992; see also the review by Mellema 1996).  The attention of researchers toward
this kind of ``secondary'' morphological component of PNe has increased
continuously since the publication of the seminal papers by Balick et al.
(1993, 1994, 1998), who also coined the term FLIERs (Fast Low Ionization
Emitting Regions) to indicate the low-ionization, small-scale structures found
in various PNe which move supersonically relative to their ambient gas.  Much
work has been done since, highlighting the existence of a large variety of
low-ionization microstructures in PNe:  some are found inside the main bodies of
the nebulae (e.g., NGC 3242; Balick et al.  1993), others outside (Fg~1; Lopez,
Roth \& Tapia 1993), some move fast (Hb 4; Hajian et al.  1997), and others
share the expansion velocity of their environments and thus are not genuine
FLIERs (e.g., NGC~7662; Balick et al.  1998); some lie along the symmetry axis
of the nebulae (NGC~7009; Balick et al.  1998), while others are apparently
directed along multiple/random directions (IC 4593, Corradi et al.  1997), etc.
In particular, several authors have focused the attention on remarkable
point-symmetrical features which strongly suggest the occurrence of episodic,
precessing ejecta from the central stars (e.g., Schwarz 1992; Corradi \& Schwarz
1993; Lopez, Meaburn \& Palmer 1993; Guerrero \& Manchado 1998).

Balick et al.  (1998) discuss the difficulties of the existing dynamical models
in accounting for all the properties of the sample of FLIERs that they observed.
While theoretical models clearly urge, further observations would help to get a
more precise characterization of all those phenomena that come under the quite
generic definition of ``low-ionization, small-scale structure'' in PNe.  In
particular, they would help in determining whether one should try to constrain
all types of low-ionization microstructures in PNe into a common, single
theoretical framework or, as it is more likely, at least to restrict their
origin and evolution to a limited number of different dynamical/radiative
processes.

In this paper, we present images and radial velocity measurements for three PNe
from the sample of Corradi et al.  (1996b), who highlighted the existence of
low-ionization small-scale structures in 23 PNe by computing 
\newline (\nii+\ha)/\oiii\
ratio maps in the image catalog by Schwarz, Corradi \& Melnick (1992).  An
analysis for other six PNe will be presented in forthcoming papers.

\section{Observations}

Images and long-slit spectra of NGC~3918 (PN G294.6+04.7), K~1--2 (PN
G253.5+10.7), and Wray 17-1 (PN G258.0-15.7) were obtained on 1996 April 26--27
at the 3.5-m~ New Technology Telescope (NTT) at ESO, La Silla (Chile), using the
EMMI multimode instrument.  With the TEK 2048$^2$ CCD ESO\#36, the spatial scale
of the instrument was 0$''$.27 ~pix$^{-1}$ both for imaging and spectroscopy.
The central wavelength and full width at half-maximum (FWHM) of the \nii\ filter
used for imaging are 658.8~nm and 3.0~nm, respectively.  Further details of the
observations are listed in Table~1.  As with the spectroscopy, EMMI was used in
the long-slit, high-resolution mode (Corradi, Mampaso \& Perinotto 1996a),
providing a reciprocal dispersion of 0.004~nm~pix$^{-1}$, and a spectral
resolving power of $\lambda$/$\Delta\lambda$=55000 with the adopted slit width
of 1$''$.0.  The slit length was of 6~arcmin.  The echelle order selected by
using a broad \ha\ filter includes the \heii\ line at $\lambda$=656.01~nm, \ha\
at $\lambda$=656.28~nm, and the \nii\ doublet at $\lambda$=654.81 and 658.34~nm.
The slit was positioned through the centers of the nebulae, along various
position angles as detailed in Table~1.

Narrow-band images of K~1-2 and Wray~17-1 were also obtained at the
1.54-m~Danish telescope at ESO La Silla on 1995 January 5--6.  With the TEK
1000$^2$ CCD ESO\#28, the spatial scale of the instrument was
0$''$.38~pix$^{-1}$.  The central wavelength and FWHM of the filters used at the
1.54~DAN were:  469.7/11.4~nm for \heii, 486.2/7.5~nm for \hb, 500.5/7.3~nm for
\oiii, 655.9/1.3~nm for \ha, 658.4/2.6~nm for \nii, and 672.7/6.3~nm for \sii.
Exposure times and seeing are given in Table~1.

We also retrieved an image of NGC~3918 from the {\it HST} archive obtained on
1995 with the WFPC2 camera (PC CCD, 0$''$.0455 pix$^{-1}$) in the F555W filter
(525.2/122.3~nm).  The emission of NGC~3918 in this broad-band {\it HST} filter
is dominated by the very strong \oiii\ lines at $\lambda=$495.9 and 500.7~nm.
This is confirmed by an \oiii\ image taken at the NTT (not presented here) which
shows the same overall structure as the {\it HST} one, apart from the obvious
coarser resolution.  However, various other nebular lines (including He $\sc ii$
468.6, \hb\, and N $\sc ii$ 575.5) fall within the transmission range of the
filter and may be important at specific positions in the nebula.

Images and spectra were reduced in a standard way using MIDAS. 

\section{NGC 3918}

NGC~3918 is a widely studied PN.  Detailed work on this object was presented by
Clegg et al.  (1987), who derived parameters for the central star and nebular
chemical abundances, and concluded that the nebula is optically thin.  From the
limited morphological and kinematical information they possessed, Clegg et al.
(1987) adopted a biconical ``pole-on'' geometry for the nebula, and derived an
expansion age of 3000~y for a distance of 1.5~kpc.  Other kinematical studies
limited to the central region of the nebula were presented by Dopita (1978) and
Pe\~na \& Torres-Peimbert (1985).

The morphology of NGC~3918 is complex, and we will focus the discussion only on
its most relevant properties, distinguishing between the large-scale components,
which determine the overall appearance of the nebula, and the low-ionization,
small-scale features, which are observed in several regions of the nebula.  The
{\it HST} and NTT \nii\ images of NGC~3918, rotated so as to have the symmetry
axis of the nebula aligned with the vertical direction, are presented in
Figure~\ref{F-ima}.

\placefigure{F-ima}

\subsection{Morphology: large-scale structures}

At high intensity levels (Fig.~\ref{F-ima}, top left), the {\it HST} image shows
a bright inner shell of roughly elliptical shape from which two fainter
protrusions extend in the vertical ($\approx $ east--west) direction, giving to
the whole structure a spindle-like appearance.  In what follows, we will call
this the ``inner shell''.  Its size is 12$''$$\times$20$''$, measured along its
minor and major axes.

It is surrounded by an elliptical ``outer shell'' of 16$''$ diameter.  At lower
intensity levels (Fig.~\ref{F-ima}, top right), the {\it HST} image shows that
the outer shell also possesses faint protrusions; they are 1.5 times more
extended than those of the inner shell.  In addition, the outer shell is
surrounded by a circular region of very faint emission showing a number of
radial filaments.  The brightness of the outer shell, as compared to the inner
one, is enhanced in \nii\ (Fig.~\ref{F-ima}, bottom left).

\subsection{Morphology: small-scale structures}

Several low-ionization small-scale struc\-tu\-res, located roughly along the major
axis of the inner shell, are clearly visible in the \nii\ image
(Fig.~\ref{F-ima}, bottom right).  For the sake of clarity, features which are
further discussed in the following are labeled in Fig.~\ref{F-ima} with the
letters $A$, $A'$, $A''$, $B$, $C$ and $C'$.

The most notable of these is the highly collimated feature on the lower
(eastern) side of the nebula which appears as a thin straight lane ($A''$)
extending out of the inner shell along its major axis (see also Corradi et al.
1996b).  At its end, this structure brightens and broadens ($A'$ and $A$).  In
the {\it HST} image, it is surrounded by a broad arc of emission, corresponding
to the southern, faintest protrusion mentioned above.

On the opposite (western) side of the central star, there is one main
condensation ($B$, first identified by Corradi et al.  1996b) lying at the edge
of the outermost upper protrusion of the nebula.  Moreover, on the same side
there is another faint, collimated bridge of emission, barely visible in our
\nii\ image and not detected previously, which extends far from the inner shell
up to positions labeled as $C'$ and $C$.  It is also aligned with the long axis
of the inner shell.

\subsection{Spatio-kinematical modeling of the inner shell}

\placefigure{F-spec}
\placefigure{F-mellema}

The NTT spectra of NGC~3918 are presented in Figure~\ref{F-spec}.  The slit was
located through the central star, along the major (P.A.=$-76$\gr) and minor
(P.A.=$+13$\gr) axes of the nebula (i.e.  vertically and horizontally,
respectively, as indicated by short lines on either side of the nebula in
Fig.~\ref{F-ima}, bottom right).  From images and spectra, we derived basic
kinematical, geometrical and orientation parameters for the spindle-shaped inner
shell of NGC~3918.

First, we note that both its shape and spectrum are consistent with the
prediction of the interacting-winds theory.  A very good qualitative fit to the
observations is in fact given by the axisymmetrical models of Mellema (1993),
which consider an initial mass distribution strongly enhanced in an
``equatorial'' plane, forming a sort of circumstellar torus with an
equatorial-to-polar density contrast as large as $\sim$10 (parameters $B$=6 and
$A$=0.9 in his calculations).  This initial mass distribution, under the action
of the expanding shock driven by the fast wind, would form a shell of compressed
gas with a spindle-like shape (Fig.~\ref{F-mellema}, left), as observed for the
inner shell of NGC~3918 (Fig.~\ref{F-ima}, top left).  The shape of the
long-axis, position--velocity plot is also qualitatively reproduced (cf.
Fig.~\ref{F-mellema}, right, with the \heii\ spectrum at P.A.=$-76$\gr\ in
Fig.~\ref{F-spec}).  According to this modeling, the protrusions of the inner
shell would develop along the {\it polar} axis of the system, the observed short
axis of the nebula being the projection on the sky of the equatorial density
enhancement.  For this reason, in the following the protrusions and jet-like
features will be referred to as ``polar'' structures.  The outer shell and its
more external extensions would in turn represent the ionized unshocked remnant
of the AGB wind.

A more detailed comparison with the model of Mellema (1993) is limited by the
fact that the figures in his article show \ha\ images and \oiii\ spectra
computed at a specific evolutionary time, while on the contrary we have \oiii\
images and \ha, \heii, and \nii\ spectra.  Detailed hydrodynamical modeling of
NGC~3918, taking into account the evolutionary status of its central star, would
be very useful in determining geometrical parameters and derive a dynamical age
for the nebula.  For the time being, we have derived the inclination and
kinematical parameters of the spindle-shaped inner shell making use of the
spatio-kinematical empirical model of Solf \& Ulrich
(1985), which allows us to treat the kind of 3-D geometry described above.  We
fit the \ha\ position--velocity plots and the shape of the inner shell with a
model in which the nebular expansion velocity $V_{\rm exp}$ increases from the
equatorial plane toward the polar axis, following the equation (Solf \& Ulrich
1985) \begin{equation} V_{\rm exp}(\phi)=V_{\rm e}+(V_{\rm p}-V_{\rm
e})\sin^\gamma(|\phi|), \end{equation} where $\phi$ is the latitude angle
(=0$^\circ$ in the equatorial plane, 90$^\circ$ in the polar directions),
$V_{\rm p}$ and $V_{\rm e}$ are the polar and equatorial velocities, and
$\gamma$ a shape parameter.  The other quantities involved in the model are the
heliocentric systemic velocity, $V_\odot$, the inclination, $i$, of the nebula
(angle between the polar axis and the line of sight), and the product \td\
containing the inseparable effects on the apparent nebular size due to the
distance, $D$, and to the kinematical age, $t$.

As with NGC~3918, we have measured by Gaussian fitting (with 2-pixel spatial
binning) the \ha\ radial velocities in the different regions of the nebula, both
at P.A.=$-76$\gr\ and P.A.=$+13$\gr.  The best fit of the shape and kinematics
of the inner shell (Fig.~\ref{F-fit}) is then obtained with the following
parameters:  $i=$75$^{+5}_{-10}$ deg; \td=1.0$^{+.5}_{-.1}$~y~pc$^{-1}$; $V_{\rm
p}$=50~\kms; $V_{\rm e}$=23~\kms; $\gamma=12$; and V$_\odot$=$-13$$\pm$4~\kms.
The systemic velocity agrees with the value $-17$$\pm$3~\kms\ measured by
Meatheringham, Wood, \& Faulkner (1988).  Adopting a distance $D$=1.5~kpc, a
kinematical age of 1500~y is computed for the inner shell, about half the value
derived by Clegg et al.  (1987).  Non-negligible deviations from the models are,
however, seen in Fig.~\ref{F-fit}, especially in the long-axis
position--velocity plot.  These deviations probably reflect the simplified
assumptions in the spatio-kinematical model (a time-independent expansion
velocity described by a convenient analytical expression).  Nevertheless, the
basic parameters relevant to the following discussion, namely the inclination,
$i$, of the nebula on the sky and the kinematical age parameter \td, are safely
constrained in the range of values quoted above, regardless of whether the shape
of the velocity field is reproduced exactly or not.  This is because $i$
basically depends only on the ``inclination'' of the kinematical plot, while
\td\ mainly depends on the apparent size of the nebula and on the polar
velocities, the former quantity being precisely measured from the images and the
latter being determined with sufficient accuracy from the modeling.

\placefigure{F-fit}

Comparison of the \ha\ velocities with the He $\sc ii$ 656.0 and \nii\ ones
shows that:  i) the expansion velocity is between 2 and 5~\kms\ lower for \heii\
than for H$\alpha$, while it is higher by a variable amount for the \nii\ ion
(this is the so-called ``Wilson effect''; Wilson 1948, 1950, 1958; see also
Dopita 1978 and Pe\~na \& Torres-Peimbert 1985 for NGC~3918); ii) the \nii\
velocity plots are much more irregular and knotty than the \ha\ and \heii\ ones
and are likely to represent the expansion velocity of the outer, rather than the
inner, shell, as suggested by the images.

The biconical model adopted by Clegg et al.  (1987) for NGC~3918 is evidently
inconsistent with the present data and the results of our modeling.

\subsection{Kinematics of the low-ionization features}

\placefigure{F-veljet}

The inclination computed for the inner shell allows us to determine the space
velocities and kinematical ages of the jet-like structures {\em on the
hypothesis that they are located along the polar axis of the inner shell}, as
strongly suggested by the observed morphology.  Their radial velocities are
presented in Figure~\ref{F-veljet}.  These are also listed in Table~2, together
with distances, deprojected velocities (adopting $i=75$\gr) and kinematical age
parameter, \td.  Radial velocities along the eastern filament increase linearly
(in absolute value) from 8 to 13 arcsec ($A''$) from the central star, and this
linear trend is also followed by the outermost emission ($A$).  Line profiles
are not symmetric but show faint extended red wings.  At position $A'$,
where the jet-like structure brightens in the images, there are also components
with substantially smaller radial velocities.  We note that the \nii/\ha\ line
ratio ``flares'' at position $A'$, being twice as large as in the inner shell
and in the rest of the jet.  Thus the articulated morphology ($A$+$A'$ and the
surrounding \oiii\ arc) and kinematics, and the peculiarly high \nii/\ha\
enhancement define a complex region of interaction at the leading edge of the
jet.

Considering now the $A'$ and $A''$ components, and deprojecting velocities using
the same inclination as for the inner shell, we find that the eastern jet
expands at considerable velocities, increasing linearly from 40~\kms\ at 8$''$
from the center to almost 100~\kms\ at its tip, $A$.  A remarkable result is
that the kinematical age of this jet is very similar to that of the inner shell.
On the other side of the nebula, the symmetry in the morphology and kinematics
shows that the very faint and extended emission on the western side of the
nebula ($C'$ and $C$) is the counterpart of the eastern jet.  Although the
signal is very faint here, also on this side there are in fact hints of a linear
increase of velocities from the end of the inner shell to the $C$ position
(Fig.~\ref{F-veljet}), and $C$ has a space velocity and kinematical age only
slightly larger than $A$.

Turning now to the bright knot $B$, this appears on the western side of the
nebula, roughly located along the major axis of the inner shell.  However, it
shows negative radial velocities indicating that it is not expanding along the
polar axis of the shell.  Its apparent association with the edge of the faintest
polar extensions of the outer nebula (see Figs.~\ref{F-ima} and \ref{F-veljet})
suggests that instead this protrusion has an orientation on the sky very
different from that of the inner shell.  Its origin is unclear.

\subsection{Discussion}

The present data allowed us to determine the basic geometrical and dynamical
properties of NGC~3918, making it an excellent test case for the theories of jet
formation in PNe.

First, the properties of the main nebula strongly resemble those expected
according to the commonly accepted scenario of the interacting-winds theories,
the basic shape and kinematics being nicely reproduced by the hydrodynamical
models in Mellema (1993).  A comparison with his calculations would imply that
the original density distribution of the stellar envelope ejected at the tip of
the AGB was markedly anisotropical, most of the mass being lost along the
equatorial plane in a toroidal configuration.  Spatio-kinematical modeling
provided us with the orientation angle on the plane of the sky, the deprojected
expansion velocities and a kinematical age for the main nebula.

Secondly, the location and kinematics of the extended, two-sided jet are fully
consistent with the hypothesis that it is expanding along the polar axis of the
original toroidal density distribution, departing approximately from the
position of the outer shock driven by the fast wind into the AGB remnant.  Using
the geometrical parameters of the main nebula, we could compute the deprojected
expansion velocities for the jet, information which is often difficult to derive
(mainly because the inclination is usually not easily determined).  The most
important properties of the jet of NGC~3918 are the following:  i) its expansion
velocity increases linearly approximately from the polar velocities of the inner
shell (40--50~\kms) to substantially larger values at its tips (100~\kms), and
ii) the kinematical age of the jet is very similar to that of the main nebula.
This latter result suggests that the axisymmetrical shell and the polar jet of
NGC~3918 were formed in the same dynamical process (slow--fast wind
interaction?), without invoking additional, later ejecta producing the highly
collimated outflow.  This indicates the existence of some focusing mechanisms
giving rise to accelerating, collimated polar flows during the dynamical
evolution of axisymmetrical shells.  As remarked by Mellema (1996), it is not
generally clear whether the polar jets and knots observed in several
axisymmetrical PNe are the result of the evolution of the main shell (i.e., of
the slow--fast wind interaction), or just another independent product (e.g.,
later ejecta) of whatever causes aspherical mass loss on the AGB.  Observations
of NGC~3918 strongly support the first hypothesis.  Along these lines, the
models by Frank, Balick, \& Livio (1996) seemed very promising since they
produce polar jets extending out of the outer shock during the interacting-winds
evolution of the main shell, starting with density distributions which are also
fully consistent with those deduced for NGC~3918.  The crucial point there, as
compared to other calculations, is to consider a gradual transition from the
slow to the fast wind, which allows the formation of a polar jet in the early
phases of the nebular evolution.  Objects with large equatorial-to-polar density
contrasts like NGC~3918 would sustain their focused flows best and make long
jets.  Dwarkadas \& Balick (1998), however, argued against this jet formation
mechanism, which would be inhibited by the presence of thin shell nonlinear
instabilities.  They also warn about the possible numerical artifacts producing
converging collimated flows in 2-D simulations.  It is hoped that the present
observations of NGC~3918, however, will stimulate further effort along the
above-mentioned lines.  For instance, the transient, jet-like protuberances
which appear in the models by Dwarkadas \& Balick (1998) at later stages of
evolution should be further considered.

Inclusion of magnetic fields offers an appealing alternative solution to the
problem of jet formation (Garc\'\i a-Segura et al.  1999).  Their
magnetohydrodynamical simulations, in fact, predict the formation of polar jets
in planetary nebulae as a consequence of sufficiently strongly magnetized
post-AGB fast winds.  A noteworthy feature of these models is that they
reproduce the linear increase of the expansion velocity along the jet that is
observed in NGC~3918.  The highly collimated outflows resulting from the models
of Garc\'\i a-Segura et al.  (1999, see also Garc\'\i a-Segura 1997) are subject
to kink instabilities, which might explain the knotty appearance of the jet of
NGC 3918.  The presence of magnetic fields would also relax the need for large
equatorial-to-polar density contrasts in the original AGB wind in order to
obtain elongated nebulae and jets.  It would be highly desirable to try to
model, along these lines, the detailed shape and kinematics of the shells and
jets of NGC~3918.

\section{K 1-2}

K~1--2 is one of the PNe known to have a close binary system at its center.  The
orbital period is of 0.676 days, which is considered to be the outcome of
shrinking of the original orbit following a common-envelope phase of the system
during the AGB evolution of the primary star (Bond \& Livio 1990).  The
existence of the jet-like feature of K~1-2 was first reported by Lutz \& Lame
(1989) and later by Bond \& Livio (1990).  Our narrow-band images are presented
in Figure~\ref{F-imak12}.

\placefigure{F-imak12}

\subsection{Morphology}

The main nebula has a low surface brightness and an elliptical shape in \ha\ and
\hb\ , while it shows extended emission (``ears'') in the EW direction in the
\oiii\ image.  Superimposed on this nebulosity, a knotty, collimated structure
is visible along P.A.=$-27$$^\circ$ in all the observed emission lines, but
particularly prominently in the light of low-ionization species like \nii\ and
\sii.  This collimated outflow consists of two strings of knots aligned radially
on opposite sides of the central star.  The central star, however, is slightly
offset from the line joining most of the features, and in addition the string of
knots shows some bending.  Some knots are resolved in our images as slightly
elongated, elliptical blobs.  On the NW side, the brightest portion of the
low-ionization structure consists of a filament (labeled $A$ in
Fig.~\ref{F-imak12}) which is aligned exactly along the radial direction from
the central star.  This filament is marginally resolved into knots in [N $\sc
ii$], and has a faint tail ($A'$) directed outwards.  A careful inspection
indicates that the brightest SE knots, albeit very small ($<2''$), show some
evidence of ionization stratification, with the \oiii\ emission slightly closer
to the central star than the \nii\ one.

In addition to this collimated system of knots, we confirm with the new imaging
presented here the discovery by Corradi et al.  (1996b) of the existence of
other low-ionization microstructures, some of them located along directions
nearly perpendicular to that of the main jet-like structure.  These other knots,
indicated by the six vertical arrows in Fig.~\ref{F-imak12}, also have a
slightly elliptical shape and form a more sparse system than the main one.

\subsection{Kinematics}

\placefigure{F-spectrak12}

The \ha\ and \nii\ long-slit spectrum of K~1-2 passing through the main system
of knots (P.A.=$-27$\gr) is shown in Fig.~\ref{F-spectrak12}.  The \ha\ spectrum
is the superposition of the emission from the jet-like features and some
extremely faint emission from the main elliptical nebula.  By binning the signal
in the central 5$''$.6 (free from any emission from the knots), a splitting of
the \ha\ line of 50~\kms\ is measured, implying an expansion velocity for the
elliptical shell of 25~\kms.  The measured heliocentric systemic velocity
$V_\odot$=65$\pm4$~\kms\ is in excellent agreement with the value of
66$\pm4$~\kms\ of Schneider et al.  (1983).  The FWHM of the \ha\ line in each
component of the elliptical shell (0.1~nm) is significantly larger than that of
expected from instrumental (0.01~nm) and thermal (0.05~nm for $T$=10000~K)
broadening.  This might reflect the existence of a velocity gradient through the
(thick) shell.

The \nii\ and \ha\ radial velocities for the system of knots were measured by
Gaussian fitting.  For \ha, velocities were measured only where it was possible
to estimate the relative contribution of the surrounding elliptical nebula.
Fig.~\ref{F-spectrak12}, right, shows the \nii\ and \ha\ velocities, which agree
with each other.  In the figure, we also plot the velocity field that would be
expected for the main elliptical shell assuming spherical symmetry, an average
shell radius of 20$''$ (measured from the \ha\ and \oiii\ images), and adopting
an expansion velocity of 25~\kms.  Although this is a crude estimate of the
kinematics in the main nebula, it is clear that all the low-ionization features
except $A'$ have velocities similar to those computed for the shell.  Unless
this comes from a conspiracy between real distances, expansion velocities, and
projection effects, this property would indicate that most of the low-ionization
features are located within the main elliptical shell and move with velocities
comparable to those of the surrounding gas (and are hence not FLIERs).  The
extended tail $A'$ of the bright NW filament $A$ has peculiar velocities
instead, larger than all the other features.  A detailed comparison of the \ha\
and \nii\ images suggests that filament $A$ is located within the main shell,
while its tail, $A'$, extends outwards; this would indicate that the gas is
accelerating as it leaves the elliptical nebula.  Under the hypotheses that i)
$A$ is indeed located within the main elliptical shell and shares its expansion
velocity, and ii) the central splitting of $\pm25$~\kms\ can be taken as a
measure of the expansion velocity of the main shell also at the location of
filament $A$, then from the observed radial velocities we obtain an inclination
on the line of sight of 40\gr\ for the collimated structure.  This number should
be taken with some caution, considering the uncertainty in the assumptions from
which it is derived.  With this inclination, space velocities in the tail $A'$
would increase with radius from 25~\kms\ to 45~\kms\ at its outermost point.

Both the \nii\ and \ha\ lines are moderately broadened, having FWHMs (not
corrected for instrumental and thermal broadening) between 0.025 and 0.045~nm,
and between 0.07 and 0.09~nm, respectively.  There is also some evidence that
the \nii\ line becomes slightly narrower in the tail $A'$ of the NW filament,
strengthening the idea that this gas is expanding freely after leaving the
shell.

No kinematical information is yet available for the sparse knots in the almost
perpendicular direction.

\subsection{Discussion}

K~1--2 consists of i) a diffuse elliptical shell expanding at a velocity of
25~\kms, ii) a main string of collimated, low-ionization knots at P.A.=$-27$\gr,
and iii) a few other knots most of which are directed in nearly perpendicular
directions.

The kinematical analysis suggests that most of the knots are embedded in the
shell and have a similar expansion velocity, apart from the extended tail on the
NW side which is composed of material accelerated to larger velocities
($\sim$45~\kms\ at its end) as it leaves the shell.  The inclination with
respect to the line of sight of the collimated outflow is estimated to be around
40\gr.  This makes knowledge of the orbital parameters of the central binary
star (especially the inclination of the orbital plane and the position angle of
the line of nodes) very important, since it would help in understanding whether
the collimated knots correspond to an ``equatorial'' (i.e., along the orbital
plane), polar, or even meridional outflow.  Some precession of the collimating
source would also need to be invoked to explain the bending of the main string
of knots.

As for the origin of this collimated feature, Soker (1997) discusses several
ways in which a close binary system such as K~1--2 can produce jets.  In his
view, polar jets are generally expected from stars that have undergone a
common-envelope phase with stellar companions.  Although this offers an
explanation for the origin of jets in PNe, their (hydro)dynamical evolution
should be worked out to allow for a detailed comparison with the present
observations.  In any case, knowledge of the spatial orientation of the jet of
K~1--2, together with future photometric, spectroscopic and polarimetric
observations of its central star, might provide the first, crucial observational
proof of the occurrence of collimated polar outflows in likely post
common-envelope central stars of PNe.  For the time being, this hypothesis is
supported by the intermediate inclination computed for the jet, since the binary
system of K 1-2 is not seen close to edge-on (it is not an eclipising binary),
nor it is seen pole-on (otherwise no significant photometric orbital variations
would be detected).

The origin of the sparse knots in the other regions of the nebula remains
unclear, lacking further spectroscopical information.

\section{Wray 17-1}

The central star of Wray 17-1 is a non-pulsating PG1159-type star.  So far there
are only about 20 hydrogen-deficient stars known (cf.  Werner 1993).  Nine of
them are surrounded by a PN, Wray 17-1 (also called Lo 3) being one of those.
Detailed non-LTE studies of various PG1159 stars have shown that they occupy a
pre-white dwarf area in the HR diagram; thus they are rather evolved objects.

A report and short description of the jet-like features of Wray~17--1 is given
by Leisy \& Dennefeld (1993).  Our narrow-band images are presented in
Figure~\ref{F-imawra171}.

\placefigure{F-imawra171}

\subsection{Morphology}

As in the case of K~1--2, at faint brightness levels the nebula appears as a
diffuse elliptical shell.  Around it, complex extended nebulosity with bright
arc-shaped limbs is observed, especially in our deep \oiii\ image.  Superimposed
on the diffuse nebula, there is a pair of regions of enhanced brightness
(labeled $A$ and $B$ in the \nii\ image of Fig.~\ref{F-imawra171}) located
symmetrically on opposite sides of the central star along the direction defined
by P.A.=$-28$\gr.  In an almost perpendicular direction (P.A.=$-103$\gr) and at
similar distances from the central star, there is a second, fainter pair of
structures composed of two small knots ($C$ and $D$) surrounded by large arcs
more visible in the \oiii\ image.  Together, the two pairs of structures form a
sort of ``four-leaf-clover'' morphology.  These four regions correspond to the
low-ionization, collimated features reported by Corradi et al.  (1996b); they
are prominent in \nii\ and [S $\sc ii$], and show a high degree of collimation
in the \nii/\oiii\ ratio map in Figure~\ref{F-imawra171}.  Each region of the
brightest pair ($AB$) looks like a diffuse patch of high-ionization gas from
which narrower low-ionization radial tails depart in the outward directions.
This is further illustrated in Fig.~\ref{F-zoom}, where we present a zoom of the
\oiii\ and \nii\ image of $A$; the \nii\ emission (and presumably also the
unobserved \oii) forms a sort of ``shadow'' of the \oiii\ patch.  These
low-ionization tails roughly follow the surface brightness of the main
elliptical nebula, disappear where the latter does, and reappear as faint outer
knots ($A'$ and $B'$ in Fig.~\ref{F-imawra171}) in correspondence to the limbs
of the extended nebulosity.

In Fig.~\ref{F-zoom}, a puzzling vertical ``bridge'' connecting $A$ to a
stellar-like object is also observed.  The bridge is real, since it persists
after careful subtraction of the point spread function of the stellar-like
object.  The bridge is most visible in the [N $\sc ii$], \sii\ and \nii/\oiii\
images.  Comparison of the brightness of the point-like object at the apex of
the bridge with that of field stars in the different narrow-band images shows
that it is very probably a red star (unfortunately it falls out of the slit used
for the kinematic measurements).

The second pair of low-ionization features at P.A=$-103$\gr\ have a different
morphology and ionization structure:  the compact knots $C$ and $D$ are located
inside two large arcs of higher ionization emission (more visible in the \oiii\
image).  The knots are observed in all emission lines, but are prominent in
low-ionization species (see also Corradi et al.  1996b).  The \nii\ knots are
marginally resolved in our NTT images, and the western one ($D$) has very faint
tails directed both inwards and outwards (see also Fig.~\ref{F-imawra171}).

Note that the central star is slightly offset from the lines joining each pair
of features.

\subsection{Kinematics}

\placefigure{F-spectrawra171}

The spectra of Wray 17-1, taken at P.A.= $-28$\gr\ and P.A.=$-103$\gr\ to include
the four low-ionization regions discussed above, are displayed in
Figure~\ref{F-spectrawra171}.  By spatially binning the signal in the central
16$''$ of the \ha\ emission at P.A.=$-103$\gr, we obtain a rough estimate of the
expansion velocity for the main nebulosity (half the line splitting) of 28~\kms\
(cf.  the value of 16~\kms\ for \oiii\ quoted in Acker et al.  1992).

We measured \nii\ and \ha\ velocities through the low-ionization features along
the two slit positions.  These are also shown in Figure~\ref{F-spectrawra171}.
Except for knot $A'$, radial velocities are smaller or similar to the adopted
expansion velocity of the main nebula, but in contrast to the case of K~1-2, it
is not possible to get reliable estimates of the projection effects because of
the complex geometry of the surrounding nebula.  Note that the arcs surrounding
$C$ and $D$ have radial velocities comparable to those of the knots (see the
\ha\ spectrum in Fig.~\ref{F-spectrawra171}), confirming the physical
association seen in the images.  Line profiles are generally symmetric, and
there is evidence for line broadening at the position of the knots $C$ and $D$,
especially in the \ha\ emission (measured FWHM=0.12~nm); this might reflect the
velocity dispersion within the arcs around the knots.

The outermost knot of the low-ionization tail of $A$, located on the limb of the
\oiii\ extended nebulosity and labeled as $A'$ in Fig.~\ref{F-imawra171}, shows
a very peculiar radial velocity ($+60$ \kms), strikingly different from that of
the inner portions of the tail ($-30$~\kms), even though they are perfectly
aligned in the images.

The adopted heliocentric systemic velocity for Wray~17--1, taken as the average
of the values derived from the \ha\ central splitting, and from the symmetry in
velocities for each pair of low-ionization structures, is V$_{\rm
sys}$=51$\pm$5~\kms.

\subsection{Discussion}

Wray 17-1 is a complex and most peculiar nebula.  The gross structure is that of
a diffuse shell, roughly circular in the inner regions but with complex extended
emission.  Embedded in the diffuse nebula are two remarkable pairs of
symmetrical, low-ionization structures.  The two pairs are located in almost
perpendicular directions (projected on the plane of the sky).  Each pair is
similar to the other in being located at comparable distances from the central
star and in having an overall ionization state that is lower than in the
surrounding nebula.  But they also show remarkable morphological differences.

One pair ($CD$) is composed of two compact, low-ionization knots surrounded by
an outward-facing higher-ionization cap.  The other pair ($AB$) presents broader
emission patches with collimated, low-ionization outward tails, whose brightness
is modulated by that of the parent nebula.  The patches are observed in all our
images in different ionization stages, thus they correspond to real density
enhancements.  The tails are seen only in low-ionization emission (\nii\ and
\sii).  To understand the nature of these tails, a key observation is the large
difference in radial velocity (90~\kms) between the inner part of tail of $A$
and its outermost knot, $A'$, on the limb of the extended nebula (the same,
albeit on smaller scale, applies to $B$ and $B'$).  It is unlikely that this
velocity difference (with sign reversal as compared to the systemic velocity of
the nebula) corresponds to a real change in velocity and direction in the same
collimated outflow to which both $A$ and $A'$ would belong.  It is more likely
that the low-ionization tail is just an ionization effect; i.e., a
lower-ionization region shielded from energetic photons from the central stars
by the inner, higher-ionization patch.  When we measure velocities along the
tail, we would then measure not the expansion velocity of an independent,
collimated outflow, but just the kinematics of the diffuse nebula, and the
velocity variations would just reflect the complex 3-D geometry, orientation and
kinematics of the overall structure.

The physical similarities (and Occam's razor) would suggest that both pairs of
low-ionization features have the same origin.  Their structural differences
might therefore reflect different initial parameters (density, size, etc.)  or a
different evolutionary stage of these condensations.  Are they, for instance,
high-density clumps which are on the way to being photo-evaporated by photons
from the central star, and observed at different evolutionary stages (cf.
Mellema et al.  1998)?  Whatever the evolution of these components, their origin
is difficult to explain.  The patches and knots appear as density enhancements
inside the main nebula.  The fact that the features in each pair are located on
opposite symmetrical positions with respect to the central star (but note the
small offset of the latter) excludes a random formation of these condensations
within the main nebula; rather, it indicates the occurrence of collimated,
two-sided ejecta (or impinging winds).  This leads to another problem, the
almost perpendicular relative orientation (projected on the sky) of the pairs.
Collimated outflows are indeed expected along the symmetry axis of aspherical
nebulae (cf.  the discussion for NGC~3918), but it is much more difficult to
think of a mechanism able to produce independent collimated outflows along
directions with have very different orientations in space (but see IC~4593;
Corradi et al.  1997).

Finally, even more mysterious is the bridge of emission connecting patch $A$ to
a star-like object which appears, in projection, inside the nebula.  The
simplest possibility is that the association is not physical, and that some
extension of the emission patch fortuitously coincides, in projection, with the
position of a background/foreground star (but note that the bridge clearly
breaks the symmetry of the emission patch which is otherwise nicely pointing
toward the central star; cf.  Fig.~\ref{F-zoom}).  The other possibility is that
the star is physically associated with the nebula, producing the bridge to the
emission patch.  In this case, the object would be a field star running through
the nebula or a companion star in a loose binary system.  No distance is
tabulated for Wray~17--1 in the literature, but for an hypothetical distance of
1~kpc, the ``companion'' star would be located at about 20000 a.u.  from the
central star of the PN and the corresponding orbital period would be longer than
a million year:  this would make it a very loose binary system indeed.  In any
case, if the star is associated with the nebula, how could it produce the
bridge?  Is it gas stripped from the emission patch by the star in its motions
through the nebula?  Or are $A$ and the bridge gas lost from the star itself
(but how to explain then the formation of the opposite symmetric patch $B$, and
of $C$ and $D$)?  With the information available it is not possible to
investigate the nature of the nebula--star connection further; this clearly
deserves future study.

\section{Conclusions}

The three PNe studied in the present article provide valuable observational
constraints in the discussion of the formation of collimated outflows and
low-ionization microstructures in PNe.

NGC 3918 proves to be a key example of the production of an extended jet along
the symmetry axis of an elongated shell.  From our data, we have determined to a
good level of accuracy the geometrical, orientation, and dynamical properties of
the shell and jet, making NGC~3918 an excellent test case to confront theories
of jet formation in PNe.  According to pure hydrodynamical calculations, the
axisymmetrical geometry of the shell would have developed starting from a very
aspherical (toroidal) mass deposition at the tip of the AGB.  Whether these
initial conditions naturally lead to the formation of the polar jet is still
matter for debate (Balick et al.  1998); the inclusion of a strong magnetic
field in the post-AGB wind (Garc\'{\i}a-Segura et al.  1999) might contribute to
a solution of the problem.

K~1--2 shows a collimated string of knots within an elliptical shell, ejected
from a close binary system which probably underwent a common-envelope phase
during the AGB phase of the star which is producing the PN.  Estimation of the
projection of these collimated structures gives us the opportunity to verify the
idea that close binaries can produce ``polar'' outflows, provided that the
orientation parameters of the binary systems are determined.  Further
photometric, spectroscopic, and/or polarimetric observations of the central
source directed toward this aim are highly desirable.

Finally, Wray~17--1 is a most unusual and interesting object, and its two pairs
of low-ionization structures located in almost perpendicular directions (and
with notable similarities/differences between each other) are hardly
reconcilable with current ideas of the production of collimated outflows in PNe.
Some of these low-ionization features (their collimated outward tails) seem to
be produced by photoionization effects.  Wray~17-1 is a case which shows that
more than one physical process should be considered in accounting for the
variety of low-ionization, small-scale structures observed in PNe.

\clearpage

\section{Acknowledgements}

We thank Garrelt Mellema for providing us with Fig.~\ref{F-mellema}, and Howard
Bond and Don Pollacco for useful discussion on the central star of K~1-2.  The
work of RLMC, EV, and AM is supported by a grant of the Spanish DGES
PB97--1435--C02--01, and that of DRG by a grant of the Brasilian Agency FAPESP
(proc 98/7502--0).

\clearpage

\figcaption[ ]{ 
     The {\it HST} and NTT images of NGC~3918, on the same scale
     and rotated so as to have the main jet-like feature along the vertical
     direction.  Both images are shown using two different intensity cuts; 
     to the left, the highest intensity levels in a linear scale; to the 
     right, the faintest structures on a logarithmic scale.  
\label{F-ima}}

\figcaption[ ]{
    The NTT long-slit spectra of NGC~3918. Images are in a logarithmic
    intensity scale, with different cuts for the \heii\,$\lambda$656.0~nm,
    \ha\,$\lambda$656.3~nm, and \nii\,$\lambda$658.3~nm lines.
\label{F-spec}}

\figcaption[ ]{
    The radiative/hydrodynamical model by Mellema (1993) which best reproduces
    NGC~3918; see text for the model parameters (figure adapted from his
    original thesis work). To the left, the model image. To the right, the
    synthetic spectrum that would be obtained through a long-slit positioned
    along the major axis of the nebula.
\label{F-mellema}}

\figcaption[ ]{
    Kinematical modeling of the inner shell of NGC 3918 using the
    description in Solf \& Ulrich (1985).  Measurements are indicated by
    dots: at the top, the heliocentric \ha\ velocities at different positions
    along the slit; at the bottom, the shape of the inner shell from the 
    {\it HST} image.  Model fits are the solid lines.
\label{F-fit}}

\figcaption[ ]{
    \nii\ velocities, corrected for the systemic velocity of the nebula,
    of the low-ionization features of NGC 3918.  Small circles are velocities
    measured with a spatial sampling of 0$''$.5, large circles are measurements
    averaged over larger areas, which refer to the total emission
    from the specific features discussed in the text (see labels on the side).
    Crosses represent velocities in the polar protrusion of the outer shell 
    toward knot $B$. The solid line is the measured \ha\ position--velocity 
    plot for the inner shell, as in Figure~\protect\ref{F-fit}.
\label{F-veljet}}

\figcaption[ ]{
    The \hb, \oiii, \ha\ and \sii\ images of K~1-2 obtained at the 1.54-m~Danish
    telescope (on a logarithmic intensity scale), and the \nii\ image taken
    at the 3.5-m~NTT  (linear scale).
\label{F-imak12}}

\figcaption[ ]{
    To the left, \ha\ and \nii\ spectra of K~1--2 along the collimated
    systems of knots at P.A.=$-27$\gr (on a logarithmic scale). To the right,
    the measured velocities, corrected for the systemic velocity of the
    nebula.  Full circles represent the \nii\ velocities, open circles the \ha\
    ones. The dashed line is the expected velocity field for the main
    elliptical nebula for pure spherical expansion with velocity 25~\kms\ and
    a radius of 20$''$ (see text).
\label{F-spectrak12}}

\figcaption[ ]{
    The images of Wray 17-1 obtained at the 1.54-m~Danish telescope, on a
    logarithmic scale. At the bottom, we also display the \nii/\oiii\ ratio
    image, in a linear scale.
\label{F-imawra171}}

\figcaption[ ]{
    A zoom of the NW patch $A$ of Wray 17-1. Note the vertical bridge,
    discussed in the text, departing from the bottom of the emission patch 
    up to the position of a stellar-like object, indicated by the arrow.
\label{F-zoom}}

\figcaption[ ]{
    To the left, the \ha\ and \nii\ spectra of Wray 17-1, on a logarithmic
    scale.  To the right, the measured radial velocities, corrected for the
    systemic velocity of the nebula. Full circles are the \nii\ velocities,
    open circles the \ha\ ones.
\label{F-spectrawra171}}

\clearpage

\begin{deluxetable}{lclc}
\tablenum{1}
\tablewidth{40pc}
\tablecaption{Log of the observations}
\tablehead{
\multicolumn{4}{c}{\bf\it Images} \\
\multicolumn{1}{l}{Object}& 
\multicolumn{1}{l}{Telescope} & 
\multicolumn{1}{l}{Filter (exposure time, min)} &
\multicolumn{1}{c}{Seeing}}

\startdata
NGC 3918  & NTT & \nii\ (2)             & 0$''$.9 \nl
	  & HST & F555W (3)             &         \nl
K 1--2    & NTT & \nii\ (5)             & 0$''$.8 \nl
          & DAN & \hb\ (20), \oiii\ (20), \ha\ (20), \sii\ (20) & 1$''$.6  \nl
Wray 17--1& NTT & \nii\ (2)        & 0$''$.9 \nl
          & DAN & \heii\ (15), \hb\ (15), \oiii\ (5), \nii\ (15), \sii\ (15)
	  & 1$''$.1 \nl
& & & \nl
\multicolumn{4}{c}{\bf\it Long--slit spectra} \nl
       &           & P.A. (exposure time, min) &        \nl
\hline \nl
NGC 3918  & NTT &  $-76$\gr\ (30), $+13$\gr\ (10) & 0$''$.9 \nl
K 1--2    & NTT &  $-27$\gr\ (60)                 & 0$''$.8 \nl
Wray 17--1& NTT &  $-28$\gr\ (60), $-103$\gr (60) &  0$''$.9 \nl
\enddata
\end{deluxetable}

\newpage
\begin{deluxetable}{lrccc}
\tablenum{2}
\tablewidth{40pc}
\tablecaption{Kinematical data for the jets and knots of NGC~3918}
\tablehead{
\multicolumn{1}{l}{} &
\multicolumn{1}{c}{d [$''$]}& 
\multicolumn{1}{c}{V$_r$ [\kms]} & 
\multicolumn{1}{c}{V$_{dep}$[\kms]} & 
\multicolumn{1}{c}{tD$^{-1}$ [yr~pc$^{-1}$]} } 

\startdata
$A$                   & --17.3 & --25  & 96  & 0.9 \nl
$A'$                  & --14.5 & --13  & 48  & 1.5 \nl
$A''$\tablenotemark{a}  & --10.3 & --13  & 51  & 1.0 \nl
Inner shell\tablenotemark{b} &    9.3 &    &  50 & 1.0 \nl
$B$                   &   13.4 & --10  &   &   \nl
$C'$\tablenotemark{c} &   24.5 & $+$21 &  80 & 1.5 \nl
$C$                   &   27.8 & $+$28 & 106 & 1.3 \nl
\enddata

\tablenotetext{a}{Middle values in the region from 8$''$ to 13$''$ 
from the centre.}
\tablenotetext{b}{The given apparent distance and deprojected 
velocity refer to the polar axis.}
\tablenotetext{c}{The velocity (age) measurements are uncertain.}
\tablecomments{$d$ is the apparent distance from the central star, V$_r$ 
the measured radial velocity, V$_{dep}$ the deprojected expansion velocity
(using $i=75^\circ$), and $tD^{-1}$ the kinematical age parameter (see
text).}

\end{deluxetable}

\end{document}